\documentclass[twocolumn]{aa} 

\usepackage{natbib}

\usepackage{graphicx}
\usepackage{dirtytalk}  
\usepackage{textcomp}
\usepackage{gensymb}    
\usepackage{wasysym}    
\usepackage{subfigure}  
\usepackage{txfonts}
\usepackage{amssymb}
\usepackage{wrapfig}
\usepackage{xcolor}

\newcommand*{\unit}[1]{\ensuremath{\mathrm{\,#1}}} 
\usepackage[]{hyperref}

\begin{document}

\title{In-flight validation of Metis Visible-light Polarimeter Coronagraph on board Solar Orbiter}

\author{A. Liberatore\inst{1} \fnmsep\thanks{\email{alessandro.liberatore@inaf.it}}
      \and S. Fineschi\inst{1} 
      \and M. Casti\inst{2} 
      \and G. Capobianco\inst{1} 
      \and L. Abbo\inst{1} 
      \and V. Andretta\inst{3}
      \and V. Da Deppo\inst{4}
      \and M. Fabi\inst{5}
      \and F. Frassati\inst{1}
      \and G. Jerse\inst{6}
      \and F. Landini\inst{1}
      \and D. Moses\inst{7}
      \and G. Naletto\inst{8} 
      \and G. Nicolini\inst{1}
      \and M. Pancrazzi\inst{1} 
      \and M. Romoli\inst{9}
      \and G. Russano\inst{3}
      \and C. Sasso\inst{3}
      \and D. Spadaro\inst{10}
      \and M. Stangalini\inst{11}
      \and R. Susino\inst{1}
      \and D. Telloni\inst{1} 
      \and L. Teriaca\inst{12}
      \and M. Uslenghi\inst{13}
      }

\institute{
         INAF - Astrophysical Observatory of Turin, Italy
     \and
        The Catholic University of America at NASA-GSFC, USA
     \and
         INAF - Astronomical Observatory of Capodimonte, Naples, Italy
     \and
         CNR - IFN, Padua, Italy
     \and
         University of Urbino Carlo Bo and INFN, Florence, Italy
     \and
         INAF - Astrophysical Observatory of Trieste, Italy
     \and
         U.S. Naval Research Laboratory, Washington, USA
     \and 
         University of Padua, Italy
     \and
         University of Florence, Italy
     \and
         INAF - Astrophysical Observatory of Catania, Italy
    \and 
        ASI - Italian Space Agency, Rome, Italy
     \and
         MPS, G\"ottingen, Germany
     \and
        INAF - IASF, Milan, Italy
         }

\date{Received -- ; accepted --}

 
\abstract
{The Metis coronagraph is one of the remote-sensing instruments of the ESA/NASA Solar Orbiter mission. Metis is aimed at the study of the solar atmosphere and solar wind by simultaneously acquiring images of the solar corona at two different wavelengths; visible-light (VL) within a band ranging from 580\unit{nm} to 640\unit{nm}, and in the HI~Ly$\alpha$ $121.6 \pm 10\unit{nm}$ ultraviolet (UV) light. The visible-light channel includes a polarimeter with electro-optically modulating Liquid Crystal Variable Retarders (LCVRs) to measure the linearly polarized brightness of the K-corona to derive the electron density.}
{In this paper, we present the first in-flight validation results of the Metis polarimetric channel together with a comparison to the on-ground calibrations. It is the validation of the first use in deep space (with hard radiation environment) of an electro-optical device: a liquid crystal-based polarimeter.}
{We used the orientation of the K-corona's linear polarization vector during the spacecraft roll maneuvers for the in-flight calibration.}
{The first in-flight validation of the Metis coronagraph on-board Solar Orbiter shows a good agreement with the on-ground measurements. It confirms the expected visible-light channel polarimetric performance. A final comparison between the first $pB$ obtained by Metis with the polarized brightness ($pB$) obtained by the space-based coronagraph LASCO and the ground-based coronagraph KCor shows the consistency of the Metis calibrated results.}
{}


\maketitle


\section{Introduction}
\label{sec:intro}
The Sun is our only opportunity to study in detail magnetically driven activity in solar-like and late main sequence stars. Moreover, the diagnostics of the physical parameters of the coronal magnetized plasma is a crucial tool for understanding phenomena that can affect the Earth magnetosphere. For this reason, many space missions aim to take data about our star. 

Solar Orbiter is one of those missions.\footnote{Solar Orbiter is the first mission of ESA Cosmic Vision 2015–2025 programme; it is an ESA-led mission with strong NASA participation. It was launched from Cape Canaveral in February 10\textsuperscript{th} 2020 at 04:03 UTC aboard a NASA-provided Atlas V 411 launch vehicle.} The scientific goal is to perform detailed measurements of the inner heliosphere and solar wind by combining observations from all the 10 on-board instruments. Solar Orbiter will allow for the first time the remote-sensing observation of the Sun from as close as 0.28\unit{AU}. Thanks to several gravitational assists with Venus and Earth, the spacecraft inclination will be increased with respect to the ecliptic plane until $\approx 30$\textdegree~allowing Solar Orbiter to observe the Sun's polar regions \citep{M_ller_2020}. Among the Solar Orbiter remote-sensing instruments, there is the Metis\footnote{In the ancient Greek mythology, Metis was the symbol of wisdom and deep thought.} coronagraph for the study of the solar corona \citep{E_Antonucci_2020, S_Fineschi_2020}.

One of the observational goals of Metis is the measurement of the linear polarization of the  Kontinuum component of the corona (K-corona). The K-corona is due to the Thomson diffusion by the coronal free electrons of the photospheric radiation and it is linearly polarized (e.g., \citeauthor{B_Inhester_2016}~\citeyear{B_Inhester_2016}, \citeauthor{NE_Raouafi_2011}~\citeyear{NE_Raouafi_2011}). The coronal emission is optically thin and contains not only the linearly polarized K-corona but also a mostly unpolarized component due to the scattering of photospheric light from dust (F-corona). Instrumental stray light would also adds up to the coronal signal. In order to eliminate these components, Metis will measure the polarized brightness ($pB$) through the acquisition of four images with different polarization angle.  A polarimeter assembly is situated along the optical path of the visible channel with the presence of an electro-optically modulating Liquid Crystal Variable Retarder (LCVR). 
From the images of the K-corona polarized brightness a map of the coronal electron density can be derived.~\citep{HC_VanDeHulst_1950} This coronal plasma quantity is essential to obtain other parameters, such as the solar wind speed, to help determining the origin and acceleration of the solar wind. 

In Section~2, we describe more in detail the Metis coronagraph and its polarimeter assembly comprising the LCVR. Section~3 presents the first in-flight validation results of the Metis LCVR. In Section~4, we report the $pB$ cross-calibration between Metis and SOHO/LASCO.

\section{Metis coronagraph}
\label{sec:metis_coronagraph}  

\begin{figure} [ht]
  \begin{center}
  \begin{tabular}{c}
  \includegraphics[width = 0.48\textwidth]{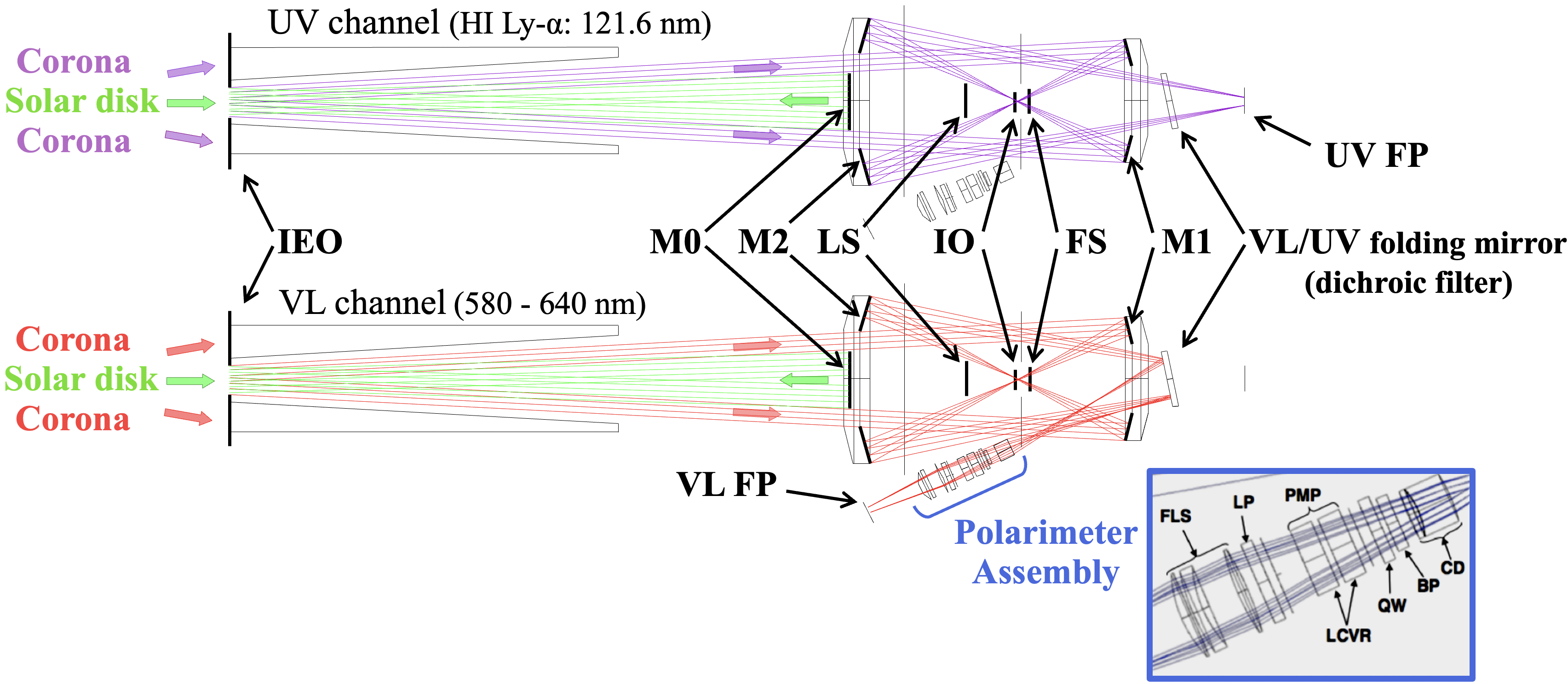}
  \end{tabular}
  \end{center}
  \caption[]{\label{fig:MetisInstrument} Metis ray trace for the UV and VL channels. The Sun-disk light is rejected by the mirror M0 while the coronal light is collected by an on-axis aplanatic Gregorian telescope. The suppression of the diffracted light off the edges of the IEO and M0 is achieved, respectively, with an internal occulter (IO) and a Lyot trap (LS). The polarimeter consists of the following elements: a collimating doublet (CD), a VL bandpass filter (BP), a focus lens system (FLS), a quarter-wave (QW) plate retarder, a polarization modulation package (PMP) comprising two liquid crystals variable retarders (LCVRs), a linear polarizer (LP) and a focusing lens system (FLS). On the focal planes (FP) are positioned the detectors.}
\end{figure}

Metis is the coronagraph on-board Solar Orbiter for the observation of the inner corona, that is, within heliocentric heights $<~9$~solar radii ($R_{\odot}$). During the mission, the spacecraft will cover a wide range of distances from the Sun. For this reason, all the instruments aboard have a particular design to deal with a high thermal flux when they are at the perihelion. In order to reduce the high thermal load on the instrument when it is near the Sun, the Metis occultation scheme was based on an \say{inverted} externally-occulted configuration. The coronal light is collected by an on-axis aplanatic Gregorian telescope while the Sun-disk light is rejected by a mirror put along the path (M0 in \figurename~\ref{fig:MetisInstrument}). The inverted external-occulter (IEO) consists of a circular aperture that reduces by two orders of magnitude the thermal load on the rejection mirror compared to that in the annular aperture of classical coronagraphs \citep{S_Fineschi_2020}.

Metis can simultaneously acquire images in visible and ultraviolet light (\figurename~\ref{fig:MetisInstrument}).  
The coronal light is splitted by a UV interferential filter that works by selecting the UV Ly$_\alpha$ (121.6\unit{nm}) in transmission and by reflecting the visible-light (VL) to the polarimeter passing through a bandpass filter at $580-640\unit{nm}$.
All the details about the optical design and performance can be found in Refs.~\citep{E_Antonucci_2020}~\citep{S_Fineschi_2020}. 

\begin{figure} [ht]
  \begin{center}
  \begin{tabular}{c}
  \includegraphics[width = 0.48\textwidth]{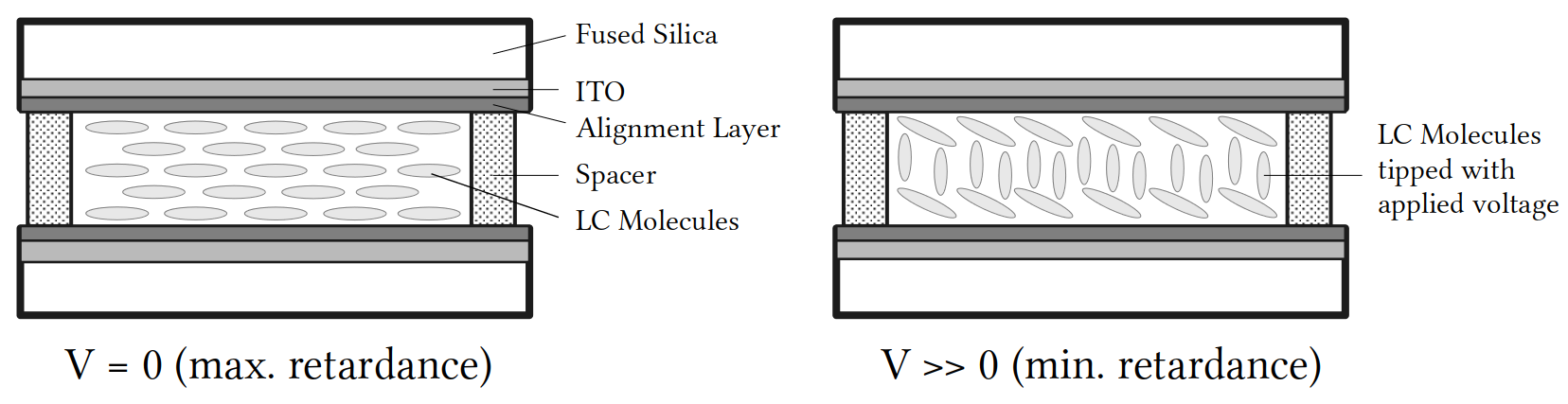}
  \end{tabular}
  \end{center}
  \caption[]{\label{fig:LCVR} 
    Schematic representation of a Liquid Crystal Variable Retarders operation.}
\end{figure}

The VL optical path comprises a polarimeter assembly with two electro-optically modulating Liquid Crystals Variable Retarders\footnote{The LCVR cells that compose the Metis PMP have been assembled with the fast axes aligned and the pre-tilt angle of the liquid crystal molecules in opposite direction, in order to obtain a wider acceptance angle (equal to $\pm$ 4\textdegree).~\citep{A_AlvarezHerrero_2011}~\citep{M_Casti_2018}} (\figurename~\ref{fig:LCVR}).~\citep{S_Fineschi_2005} \citep{L_Zangrilli_2009} An LCVR consists of optically anisotropic liquid crystal molecules embedded between two glasses with a conductive film with an ordered orientation. They have an effective birefringence value that can be changed by applying an electric field to the cells that rotates the molecules. LCVRs technology has many advantages and this is the first time that is used in a space mission.\footnote{It is used also in the Polarimetric and Helioseismic Imager (PHI) instrument on-board Solar Orbiter \citep{Solanki_2020}.} Retarders based on liquid crystals do not use moving mechanical parts, reducing noise, failure probability, and mass. The power consumption of the liquid crystal-based retarders is reduced and their response is very fast (of the order of milliseconds), providing in this way a fast modulation of the polarization state of light.

\begin{figure} [ht]
  \begin{center}
  \begin{tabular}{c}
  \includegraphics[width = 0.45\textwidth]{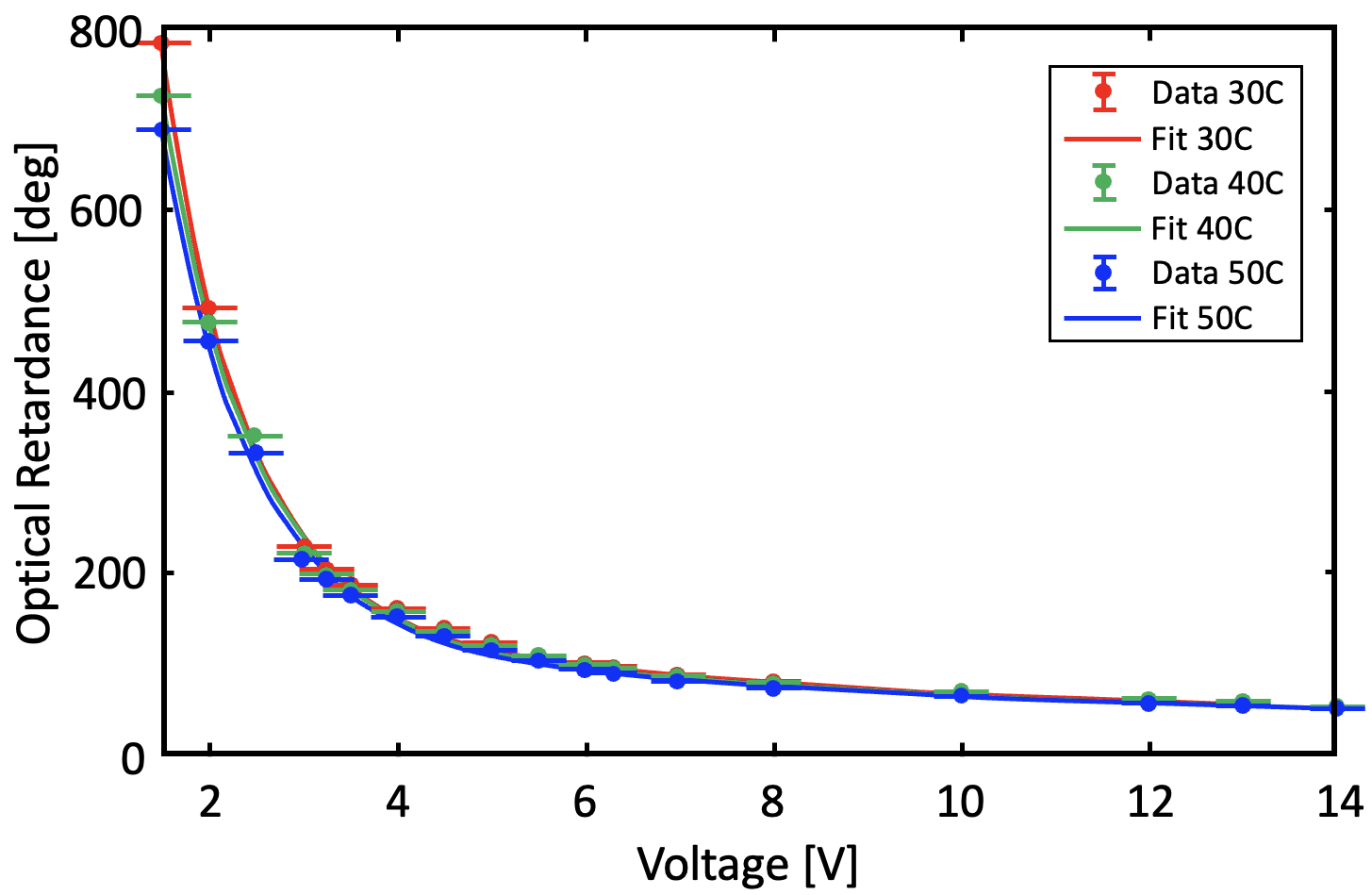}
  \end{tabular}
  \end{center}
  \caption[]{\label{fig:differentTandV} 
    Optical retardances for different applied voltages and verification of the PMP performance for different temperatures.~\citep{M_Casti_2018}}
\end{figure}

LCVR devices are sensitive to temperature (\figurename~\ref{fig:differentTandV}). Indeed, the retardance depends on the temperature of the liquid crystals cells because the different temperatures can facilitate (if higher) or contrast (if lower) the rotation of the molecules. For this reason, Metis Polarization Modulation Package (PMP) has a temperature controller to guarantee the temperature stability on liquid crystal cells. Nominally, the LCVR works at a temperature of 30\degree C. However, a characterization of the LCVR performance at different temperatures was also performed during the on-ground calibrations and can be found in \citep{M_Casti_2018}.

The on-ground calibration yielded the Metis polarimeter demodulation tensor \textbf{X}$^\dag$. 
This tensor is obtained by computing and inverting the modulation matrix \textbf{X} associated with each pixel of the acquired polarimetric images. The elements of these demodulation matrices are collected in images, each relating to an element of the matrix (obtaining a tensor of 12 images, considering that the demodulation matrices have dimension $3\times4$, see Eq.~\ref{eq:X_demod_matrix_theo}). 

The demodulation tensor polarimetrically characterises the incoming light by returning its Stokes vector \textbf{S} $= (S_0, S_1, S_2) = (I, Q, U)$.\footnote{The K-corona is linearly polarized due to the Thomson scattering of the solar disk radiation by the free electrons in the coronal medium. Being the fourth Stokes parameter ($S_3 \equiv V$) associated with circular polarization, we can consider just the first three Stokes parameters $(S_0, S_1, S_2) \equiv (I, Q, U)$ and set $V = 0$.} In accordance with the considered Polarimeter Instrument Level System (PILS) reference france PILS\footnote{X$_{PILS}$ = parallel to the incoming light direction, pointing the VL detector; Z$_{PILS}$ = parallel to the Metis linear polarise acceptance axis; Y$_{PILS}$ = to complete the right handed reference system.} (\figurename~\ref{fig:LCVR_calib_different_mi_considered_regions}) we can define the these Stokes parameters as:

\begin{itemize}
    \item[] $I$ $\equiv$ Intensity of the linearly polarized radiation beam along $\hat{n}_{Z_{PILS}}$ plus
    \item[]\hspace{6mm} intensity of linearly polarized radiation along $\hat{n}_{Y_{PILS}}$:
    \item[]\hspace{6mm} $I_{\hat{n}_{Z_{PILS}}} + I_{\hat{n}_{Y_{PILS}}}$\\
    
    \item[] $Q$ $\equiv$ Intensity of linearly polarized radiation along $\hat{n}_{Z_{PILS}}$ minus intensity
    \item[]\hspace{7mm} of linearly polarized radiation along $\hat{n}_{Y_{PILS}}$:
    \item[]\hspace{7mm} $I_{\hat{n}_{Z_{PILS}}} - I_{\hat{n}_{Y_{PILS}}}$\\
    
    \item[] $U$ $\equiv$ Intensity of linearly polarized radiation along $(\hat{n}_{Z_{PILS}} + \hat{n}_{-Y_{PILS}})/\sqrt{2}$ 
    \item[]\hspace{7mm} minus intensity of linearly polarized radiation along $(\hat{n}_{Y_{PILS}} + \hat{n}_{Z_{PILS}})/\sqrt{2}$:
    \item[]\hspace{7mm} $I_{(\hat{n}_{Z_{PILS}} + \hat{n}_{-Y_{PILS}})/\sqrt{2}} - I_{(\hat{n}_{Y_{PILS}} + \hat{n}_{Z_{PILS}})/\sqrt{2}}$\\
\end{itemize}

\noindent where $\hat{n}_{XPILS}$, $\hat{n}_{YPILS}$, and $\hat{n}_{ZPILS}$ are the unit vectors parallel to the $x$, $y$, and $z$-axis in the PILS reference frame. From the Stokes parameters it is possible to define the polarized brightness:

\begin{equation}
\label{eq:pB_def}
    pB = \sqrt{Q^2 + U^2}
\end{equation}

\begin{figure} [ht]
  \begin{center}
  \begin{tabular}{c}
      \includegraphics[width = 0.48\textwidth]{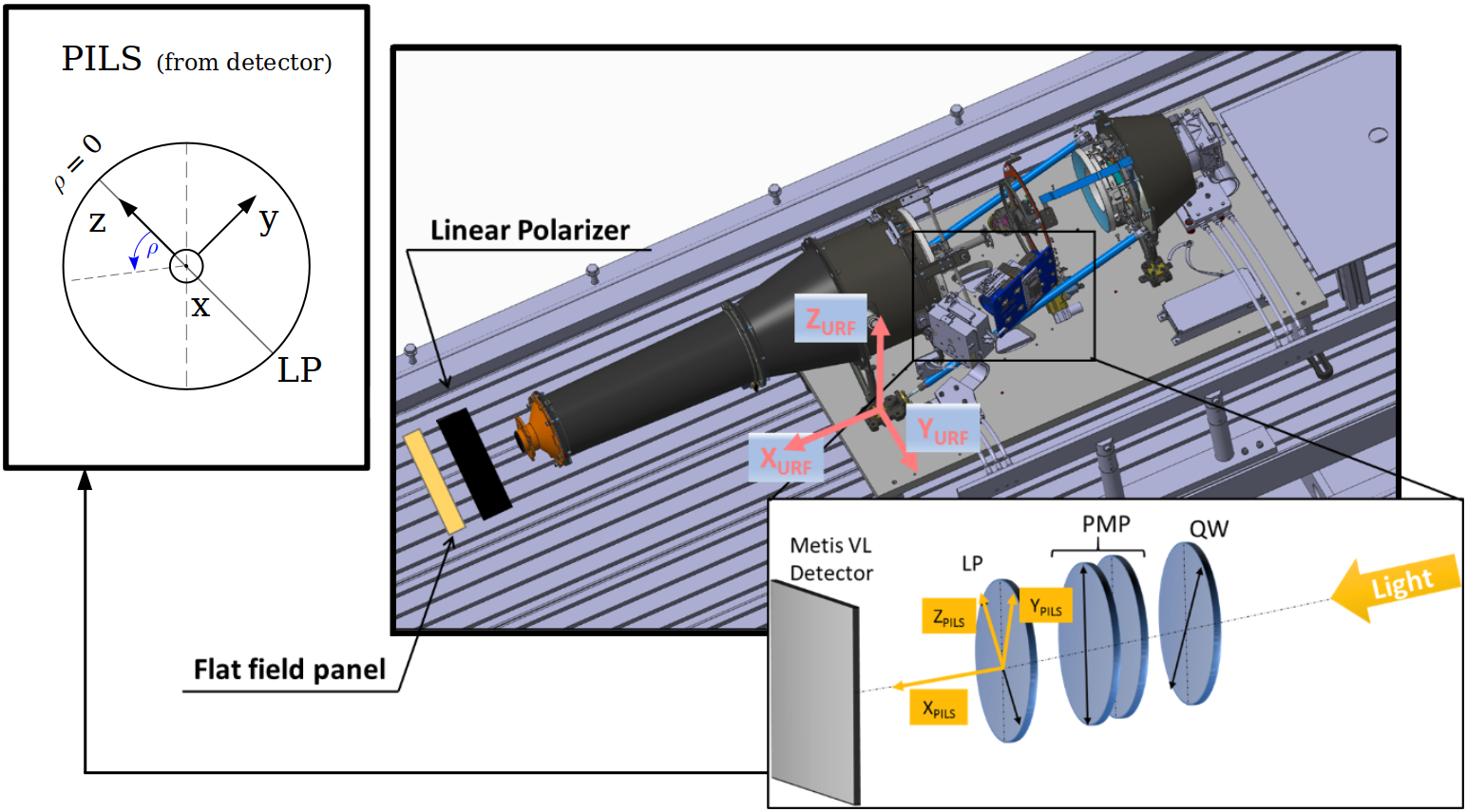}
  \end{tabular}
  \end{center}
  \caption[]{\label{fig:RS_Metis_ongroundClib}
    Schematic view of the set up used for data acquisition and Metis polarimeter optical elements relative orientations in PILS reference frame [compared with the Unit Reference Frame (URF)].
    The arrows in the optical elements represent, respectively: the fast axes for both the QW and the LP, and the PMP polarization axis.}
\end{figure}

Four images \textbf{m}~$= [m_1, m_2, m_3, m_4]$ at different retardance values are obtained by varying the voltage applied to the LCVRs and then they are combined through the relation $\textbf{m} = \textbf{X} \textbf{S} \quad\rightarrow\quad \textbf{S} = \textbf{X}^\dag \textbf{m}$. Where the theoretical modulation matrix is given by Eq.~\ref{eq:X_mod_matrix} as follow:

\begin{equation}
\label{eq:X_mod_matrix}
    \textbf{X} = \frac{1}{2}
    \begin{pmatrix}
        1 & \cos{\delta_1} & \sin{\delta_1} \\
        1 & \cos{\delta_2} & \sin{\delta_2} \\
        1 & \cos{\delta_3} & \sin{\delta_3} \\
        1 & \cos{\delta_4} & \sin{\delta_4} \\
    \end{pmatrix}
    = \frac{1}{2}
    \begin{pmatrix}
        1 & \textcolor{white}{+}0  & \textcolor{white}{+}1  \\
        1 & -1  & \textcolor{white}{+}0  \\
        1 & \textcolor{white}{+}0 & -1  \\
        1 & \textcolor{white}{+}1  & \textcolor{white}{+}0 \\
    \end{pmatrix}
\end{equation}

where the values of the quadruplet retardances, $\delta_i,\ i=1, 2, 3, 4$, of the LCVR in the polarimeter are $\delta_1=90\degree,\ \delta_2=180\degree,\ \delta_3=270\degree,\ \delta_4=0\degree$.
\noindent The theoretical demodulation matrix \textbf{X}$^\dag$ (Eq.~\ref{eq:X_demod_matrix_theo}) is obtained as the Moore-Penrose inverse of the theoretical modulation matrix \textbf{X}.

\begin{equation}
\label{eq:X_demod_matrix_theo}
    \textbf{X}^\dag = \frac{1}{2}
    \begin{pmatrix}
        1  & \textcolor{white}{+}1  &  \textcolor{white}{+}1 &  \textcolor{white}{+}1 \\
        \textcolor{white}{+}0  & -2  & \textcolor{white}{+}0 &  \textcolor{white}{+}2 \\
        \textcolor{white}{+}2  & \textcolor{white}{+}0  &  -2 & \textcolor{white}{+}0 \\
    \end{pmatrix}
\end{equation}

As mentioned, the experimental demodulation tensor associated with the Metis polarimeter was derived during the on-ground calibration by acquiring images of a known polarized source modulated by applying different voltages to the LCVR cell. By considering the elements of the Stokes vectors associated to the known polarized source (generated by a rotating linear pre-polarizer), a linear system was solved for each detector pixel to derive the demodulation tensor \citep{M_Casti_2018, Liberatore_2021_SPIE}. More information about the on-ground calibration of the Metis VL polarimetric channel can be found in Refs.~\citep{M_Casti_2019},~\citep{S_Fineschi_2020}.

\section{Polarimeter in-flight validation}
\label{sec:LCVR_in-flight_calibration}  
In this section we analyze the in-flight validation of the Metis polarimeter carried out during the first image acquisitions by the Metis coronagraph.

\subsection{LCVR retardances evaluation}
\label{subsec:InFlight_LCVR_calibration}
The K-corona is due to the Thomson diffusion of the photospheric radiation by the coronal free electrons in the solar atmosphere. It is linearly polarized and the polarization vector is tangent to the solar limb. These properties of the K-corona polarization vector can be used to perform the in-flight validation of the LCVR retardances. 

\begin{figure}[ht!]
    \centering
    \includegraphics[width = 0.78\linewidth]{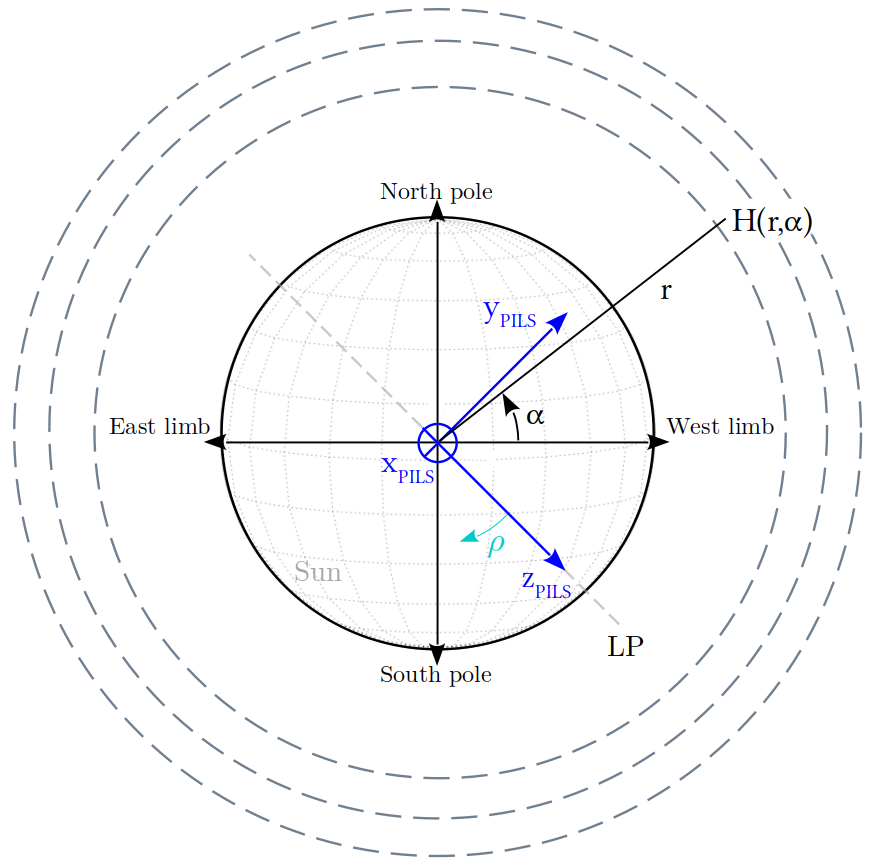}
    \caption[Polar angle at different heliocentric distances]{Positions on the sensors elements, $H(r, \alpha)$, for the LCVR retardances evaluation in Eq.~\ref{eq:3param_regression}. We considered different heliocentric distances, $r$, from $3.14~\unit{R_\odot}$ to $3.63~\unit{R_\odot}$ and different polar angles, $\alpha$. The PILS reference frame position relative to the Sun is shown in blue (in particular, the $z_{PILS}$ axis results to be at -45\textdegree~from the West solar limb).}
    \label{fig:LCVR_calib_different_mi_considered_regions}
\end{figure}

\noindent The K-corona polarization vector crossing the transmission axis of the polarimeter analyzer with an angle, the "polar angle", $\alpha$, shown in \figurename~\ref{fig:LCVR_calib_different_mi_considered_regions}, results in an intensity modulation in each single $i-th$ image of the set of 4 coronal images, acquired at different voltages of the  polarimeter. This modulation is calculated from the recorded signal, $m_i$, for each sensor element $P(r, \alpha)$ detrended of the total intensity, $I^{(0)}$, and normalized by the $pB^{(0)}$ \citep{D_Elmore_2000}:

\begin{equation}
\label{eq:2m_minus_I_over_pB}
	\frac{2m_i - I^{(0)}}{pB^{(0)}} = \cos[2(\alpha - \rho_i)]
\end{equation} 

\noindent where $m_i$ are the in-flight data and the angles $\rho_i$ are equivalent to effective polariser rotations equal to half the LCVR's retardances, $\rho = \delta/2$. The initial evaluations of $I^{(0)}$ and $pB^{(0)}$ are carried out by using the ground calibrated effective polariser rotation angles, $\rho_i^{(0)} = \delta_i^{(0)}/2$. The in-flight retardance values, $\delta_i$, are retrieved through the following 3-parameters (P0, P1, P2) regression: 

\begin{equation}
\label{eq:3param_regression}
    y_i = P_0 + P_1 \cos{[2(x - P_2)]}\\[6pt]
\end{equation} 

\noindent where $y_i = (2m_i - I)/pB$, $x$ is the polar angle $\alpha$, $P_0$ is the bias of fitting curve, and $P_1$ is the modulation amplitude. The $P_2$ parameter is defined as:

\begin{equation}
\label{eq:3param_regression_P2}
   P_2 = \rho_i - 45\textrm{\textdegree}\\[6pt]
\end{equation} 

\noindent where the shift of $-45$\textdegree~is introduced to align the PILS reference frame with the solar coordinate system (aligning the $z_{PILS}$-axis with the solar East-West direction; \figurename~\ref{fig:LCVR_calib_different_mi_considered_regions}).

\begin{figure} [ht!]
    \begin{center}
    \begin{tabular}{c}
    \includegraphics[width = 0.48\textwidth]{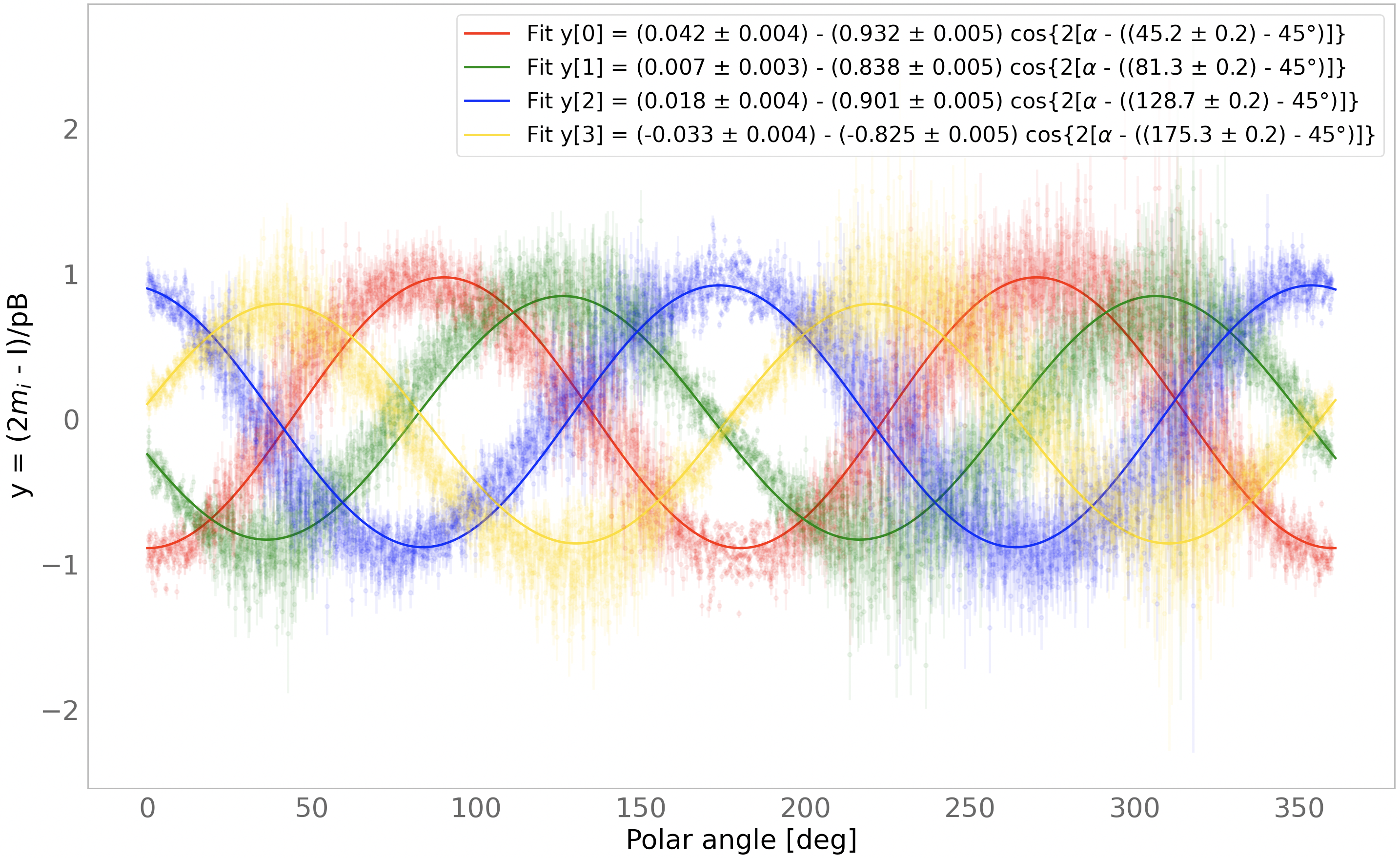}
    \end{tabular}
    \end{center}
    \caption[]{\label{fig:LCVR_calib_different_mi} 
    Intensity modulation in each $i-th$ image of the polarization set of 4 coronal images, expressed in Eq.~\ref{eq:2m_minus_I_over_pB}, as a function  of the "polar angle", $\alpha$, shown in \figurename~\ref{fig:LCVR_calib_different_mi_considered_regions}. The LCVR retardances, $\delta_i$, were derived from the regression in Eq.~\ref{eq:3param_regression}. The data come from a quadruplet acquired during the \say{roll n.0} of the IT-7 campaign (on June 8$^{th}$, 2020 - \figurename~\ref{fig:DoLP_Roll}) for a fixed heliocentric height (620 pixels from the Sun center in this particular case).}
\end{figure}

A regression is calculated for each $y_i$ obtained from the respective image of the quadruplet $m_i$, for a fixed distance from the center of the Sun as obtained by astrometry (\figurename~\ref{fig:LCVR_calib_different_mi}).

The $P_2$ parameters give the polarization angles $\vartheta$ for different heliocentric heights. We considered 10 heliocentric distances from $3.14\unit{R_\odot}$ to $3.63\unit{R_\odot}$ from the Sun center (SC) in order to avoid noisy regions at the inner and outer edges of the field-of-view (\figurename~\ref{fig:LCVR_calib_different_mi_considered_regions}). The expected average values from ground calibration for the nominal quadruplet polarization, $\rho_i$, are: (49.1\degree, 84.3\degree, 133.2\degree, 181.1\degree) with an pixel-by-pixel, flat-field uncertainty $\pm 5\degree$~each (Q4 in \tablename~\ref{tab:quadruplets_details}). From the average of the values at different heliocentric heights, we obtain: ($\rho_1=45.0\degree \pm 0.1\degree$, $\rho_2=81.4\degree \pm 0.1\degree$, $\rho_3=128.7\degree \pm 0.1\degree$, $\rho_4=175.4\degree \pm 0.1\degree$). The errors are obtained from the sinusoidal fits (\ref{fig:LCVR_calib_different_mi}).

The values of the retardances of the LCVR measured in-flight are within the uncertainty, due to the flat field, in the retardances derived on-ground. 

Even if these values have a lower error than those obtained during the on ground calibration, it was decided to continue to use the latter ones because they were obtained through a pixel by pixel process (through the use of a demodulation tensor). On the other hand, the in-flight ones were evaluated only for some fixed heliocentric heights and polar angles. However, the obtained result shows the consistency between the on-ground and the in-flight results. It is the first in-flight validation of a polarimeter with liquid crystal-based polarimeter. It may be useful to repeat this process in the future to check the status of the polarimeter over time.


\subsection{Different voltage configurations}
\label{subsubsec:different_quadruplets}
During the first remote sensing checkout window (RSCW1 - June 18$^{th}$, 2020 - distance to the Sun:  0.52 A.U.), we acquired polarimetric sequences with different quadruplets of PMP voltages. A quadruplet consists of 4 images at 4 LVCR voltages (i.e., effective \say{polarization angles}) separated by $\approx 45\degree$ from each other. The ground calibration yielded the demodulation tensor associated to each of these quadruplets (in addition to the nominal one). The goal of these different acquisitions is to check the polarimeter's response for the other set of 4 voltage configurations. \tablename~\ref{tab:quadruplets_details} summarises the considered quadruplets, with an effective angle error of $\approx1\degree$.

\begin{figure} [ht]
    \begin{center}
        \begin{tabular}{c}
            \includegraphics[width = 0.96\linewidth]{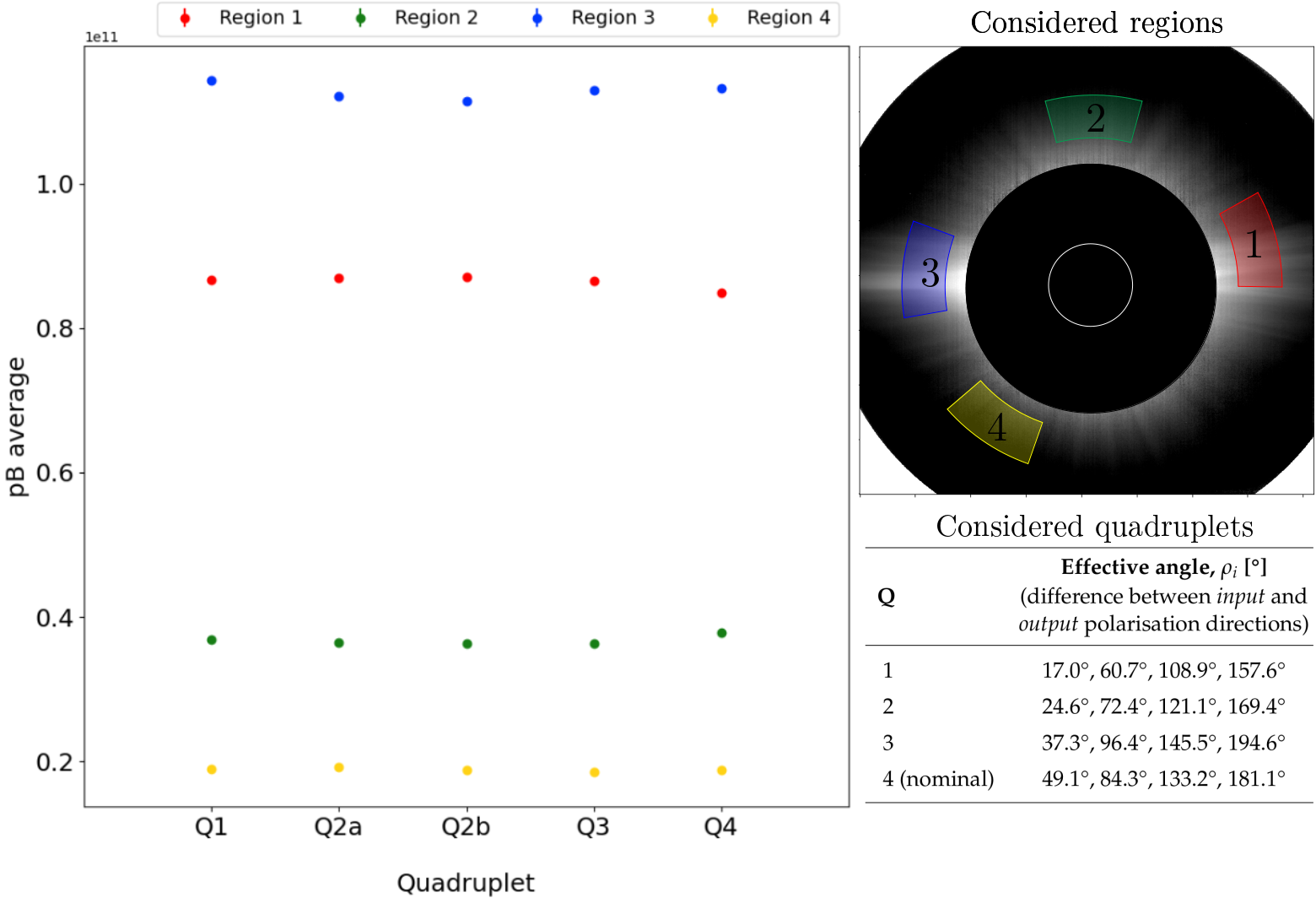}
        \end{tabular}
    \end{center}
    \caption[]{\label{fig:quadruplets_comparison} 
    Comparison between the average of the four $pB$ regions for the different quadruplets Q$_i$ (see \tablename~\ref{tab:quadruplets_details}). The differences between the different $pB$ are less than 2\%.}
\end{figure}

The demodulation tensor associated with each considered quadruplet returns the corresponding $pB^{(1)}$, that are refined values with respect to the initial $pB^{(0)}$ obtained from the ground-calibrated quadruplets\footnote{In particular we have 5 $pB$ because we performed two acquisitions with \say{quadruplet 4}.}. In the following, these refined values of polarized brightness will be indicated as $pB \equiv pB^{(1)}$, for short. Considering 4 different coronal regions (near the streamers -Region 1, 3- and near coronal holes -Region 2, 4-) and performing an average over the pixels inside these regions (\figurename~\ref{fig:quadruplets_comparison}), we can compare the measured $pB$ with different voltage configurations. The resulting $pB$ should be the same for each configuration. \figurename~\ref{fig:quadruplets_comparison} shows that the differences between the $pB$ returned by different quadruplets are $\le 2\%$. 

\section{Validation during spacecraft roll}
\label{sec:calib_during_roll}

\begin{figure} [ht]
    \begin{center}
    \begin{tabular}{c}
    \includegraphics[width = 0.48\textwidth]{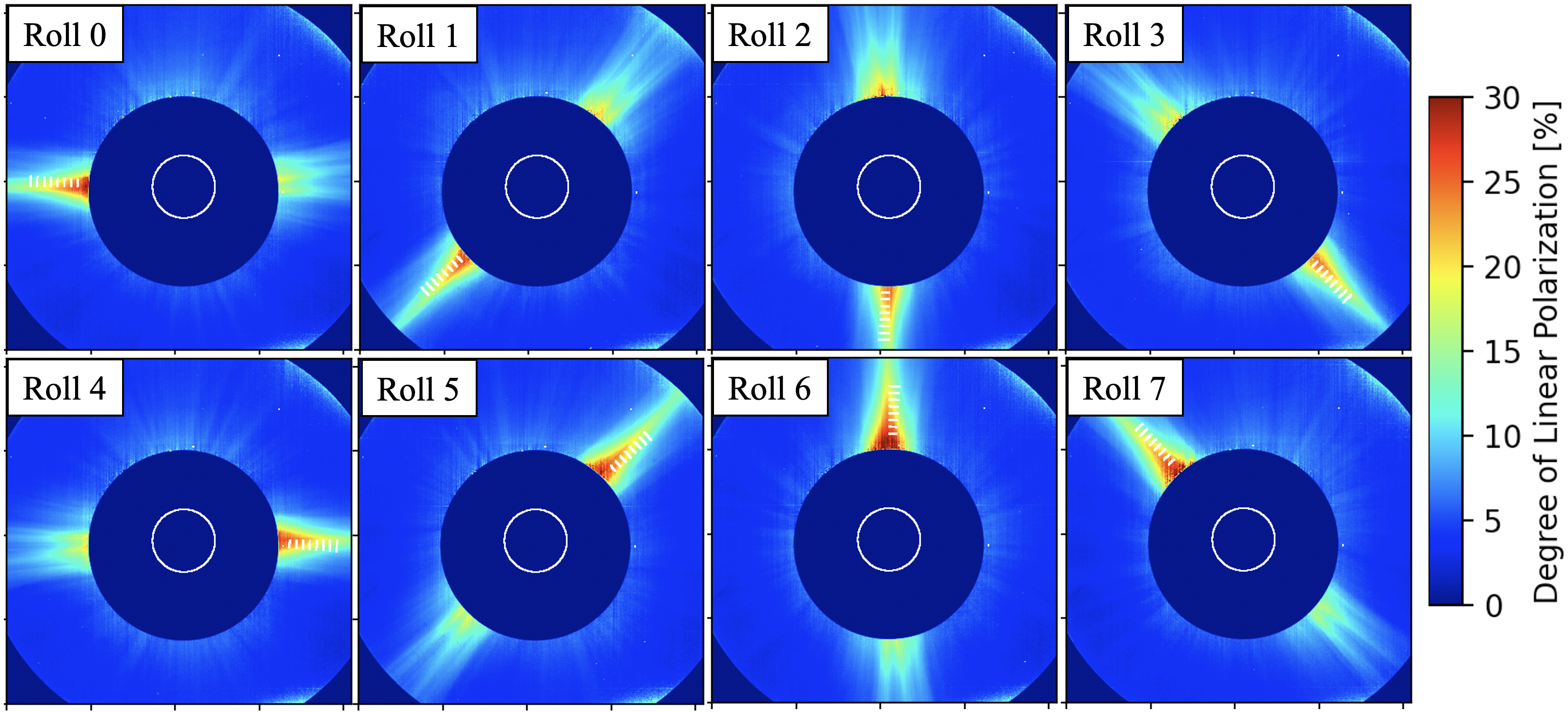}
    \end{tabular}
    \end{center}
    \caption[]{\label{fig:DoLP_Roll} 
    Degree of linear polarization ($pB$/$I$) of the K-corona for the eight different S/C roll positions and considered region (white rectangles) to evaluate the polarimetric flat field goodness. The white circle inside the internal occulter show the Sun size and position.}
\end{figure}

We acquired different image sets during a complete roll performed by the spacecraft on June 8$^{th}$, 2020 (IT-7 campaign; S/C - Sun distance: 0.52\unit{A.U.}). During the roll maneuver, the Metis polarimeter acquired a total of 8 K-corona $pB$ images (one for each roll position - see \figurename~\ref{fig:DoLP_Roll}). The comparison among these images gives the differences in the polarimetric response at different positions across the polarimeter and the detector.

The Degree of Linear Polarization (DoLP) is derived from the ratio between $pB$ and the first Stokes parameter $I$. \figurename~\ref{fig:DoLP_Roll} shows the $pB$/$I$ images for different spacecraft roll positions. For a few selected coronal regions on the frame (for example along a streamer), at  different heliocentric distances (the Sun's center position behind the internal occulter is derived from astrometry measurements), the $pB$/$I$ is plotted as a function of roll angle. 

\subsection{Check on the LCVR retardances during S/C roll}
\label{subsec:LCVR_check_roll}

\begin{figure} [ht!]
    \begin{center}
    \begin{tabular}{c}
    \includegraphics[width = 0.48\textwidth]{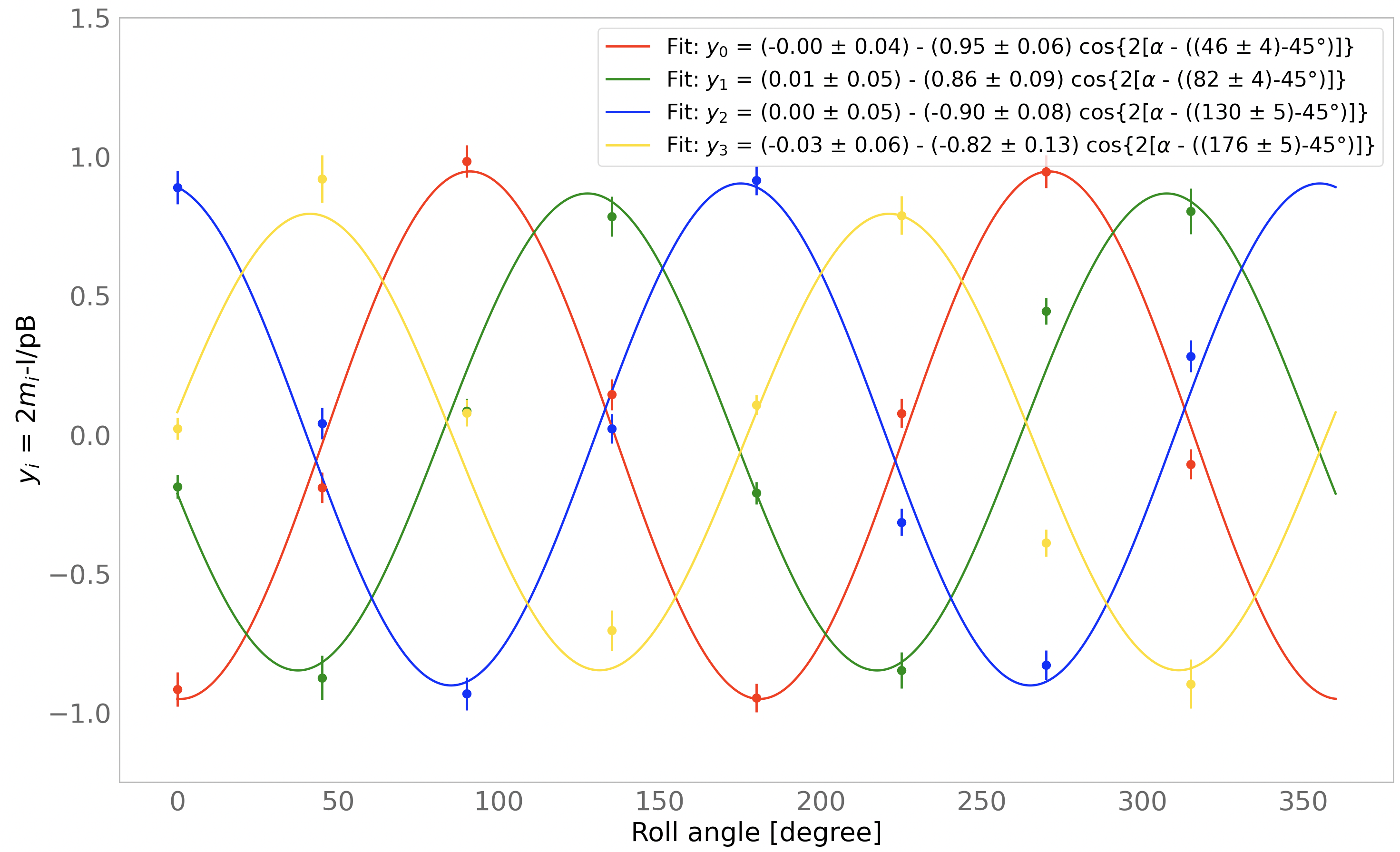}
    \end{tabular}
    \end{center}
    \caption[]{\label{fig:Malus_Roll} 
    Malus curve obtained from $y_i = (2m_i-I)/pB$ average values over the pixels in a fixed region in the solar corona during each roll.}
\end{figure}

By analizing the data acquired during the spacecraft rolls, we can repeat the same type of analysis performed in Subsec.~\ref{subsec:InFlight_LCVR_calibration}. For a given region in the solar corona, the spacecraft roll acts as a \say{rotating polarizer}. Thefore, the recorded intensity of a selected region follows the Malus's law (\figurename~\ref{fig:Malus_Roll}). The fitting parameters of this function give an estimation of the LCVR retardances that can be compared with those measured from the ground calibration and with those obtained from the previous method. 

\tablename~\ref{tab:LCVR_values_comparison} summarizes the retardances measured with the on-ground calibration, with the Malus curve fitting  Eq.~\ref{eq:2m_minus_I_over_pB} (cfr., \figurename~\ref{fig:LCVR_calib_different_mi}), and with the Malus curve obtained from the roll maneuver (cfr., \figurename~\ref{fig:Malus_Roll}).

\begin{table}[ht]
  \caption[]{Comparison between the LCVR polarization angles from on-ground calibration and in-flight validation.} 
    \label{tab:LCVR_values_comparison}
    \begin{center}       
        \begin{tabular}{cccc}
        \noalign{\smallskip}\hline
        \rule[-1ex]{0pt}{3.5ex} 
        & & \textbf{In-flight}  & \textbf{In-flight} \\
        \textbf{Theo.}& \textbf{On-ground} & (polar angle, & (roll maneuver, \\
        & & \figurename~\ref{fig:LCVR_calib_different_mi}) & \figurename~\ref{fig:Malus_Roll}) \\
        \noalign{\smallskip}\hline\noalign{\smallskip}
        \rule[-1ex]{0pt}{3.5ex}
        45 & 49\degree $\pm$ 5\degree & $45.0\degree \pm 0.1\degree$ & $46\degree \pm 4\degree$ \\
        \rule[-1ex]{0pt}{3.5ex}
        90 & 84\degree $\pm$ 5\degree & $81.4\degree \pm 0.1\degree$ & $82\degree \pm 4\degree$ \\
        \rule[-1ex]{0pt}{3.5ex}
        135 & 133\degree $\pm$ 5\degree & $128.7\degree \pm 0.1\degree$ & $130\degree \pm 5\degree$ \\
        \rule[-1ex]{0pt}{3.5ex}
        180 & 181\degree $\pm$ 5\degree & $175.4\degree \pm 0.1\degree$ & $176\degree \pm 5\degree$ \\
        \noalign{\smallskip}\hline
        \end{tabular}
    \end{center}
\end{table}

\subsection{Polarized flat field verification}
\label{subsec:PolarizedFlatField}

We verified in flight the polarized flat field by considering the $pB$ of coronal structures measured as they moved to different locations across the detector, during  roll maneuvers of the Solar Orbiter spacecraft as shown in \figurename~\ref{fig:DoLP_Roll}. 

\begin{figure} [ht!]
    \begin{center}
    \begin{tabular}{c}
    \includegraphics[width = 0.48\textwidth]{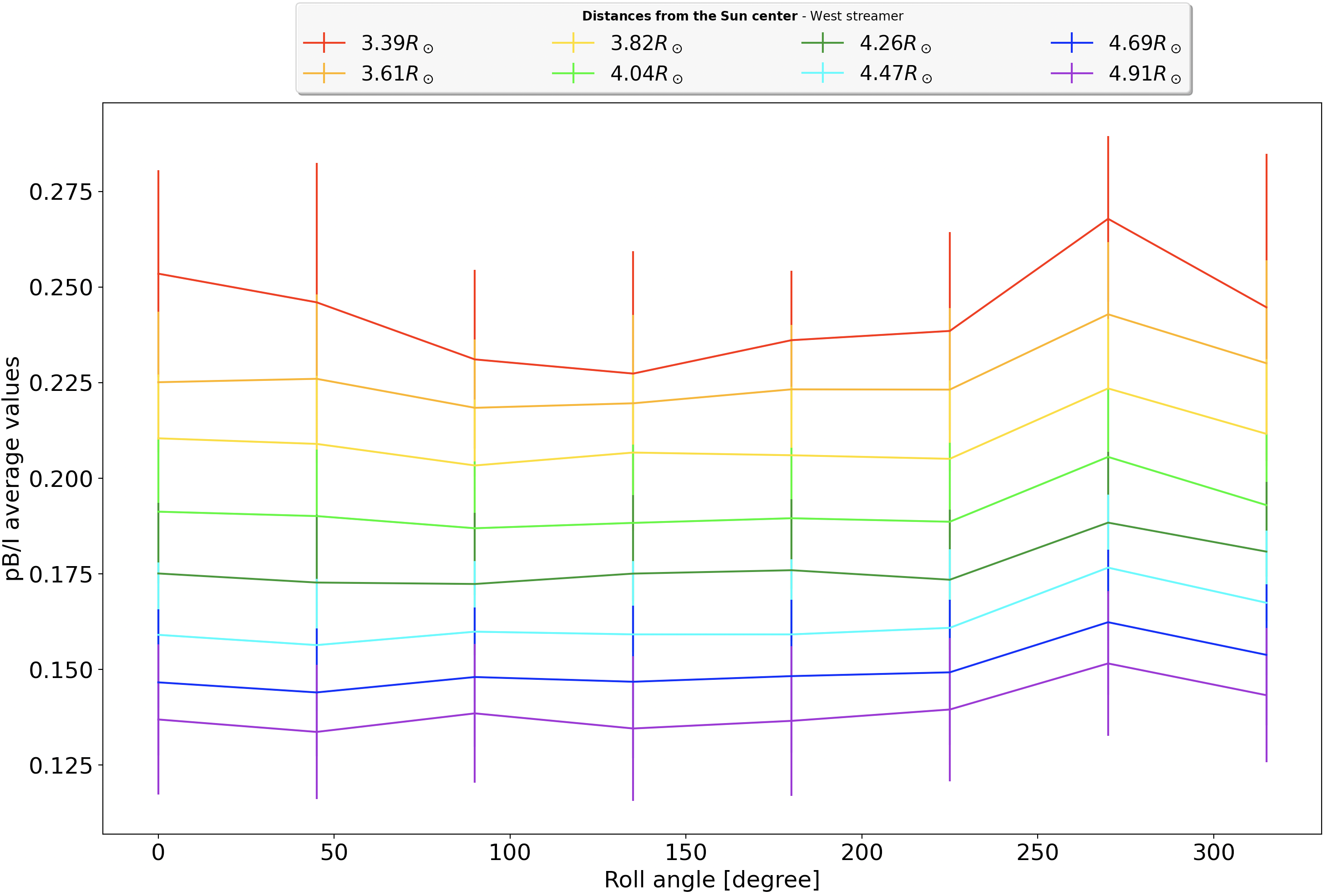}
    \end{tabular}
    \end{center}
    \caption[]{\label{fig:DoLP_Roll_values_XLab} 
    $pB$/$I$ average over the pixels inside the selected regions (for each roll) by using the demodulation tensor obtained during the ground calibration. The bars represent the intensity uncertainty inside the considered area.}
\end{figure}

\figurename~\ref{fig:DoLP_Roll_values_XLab} shows, as a function of roll angle, the $pB$/$I$ derived from the demodulation tensor, $X^\dag$, measured during the on-ground calibrations.
The quality of the polarization flat-fielding obtained by applying the on-ground calibrated demodulation tensor is indicated by the constant $pB$/$I$ values (i.e., $< 5\%$; except for the  heliocentric height 3.39 R$_\odot$, with percentage variation $\sim 12\%$), for different roll angles, the same heliocentric height. The degree of linear polarization decreases for increasing heliocentric distances: from $(0.24 \pm 0.01)$ at $3.39\unit{R_\odot}$ to $(0.14 \pm 0.01)$ at $4.91\unit{R_\odot}$. Is it possible to notice that, at higher heliocentric heights, the behavior of pB/I during the roll maneuvers results to be less flat due to the decrease of the S/N.

\begin{figure} [ht!]
    \begin{center}
    \begin{tabular}{c}
    \includegraphics[width = 0.48\textwidth]{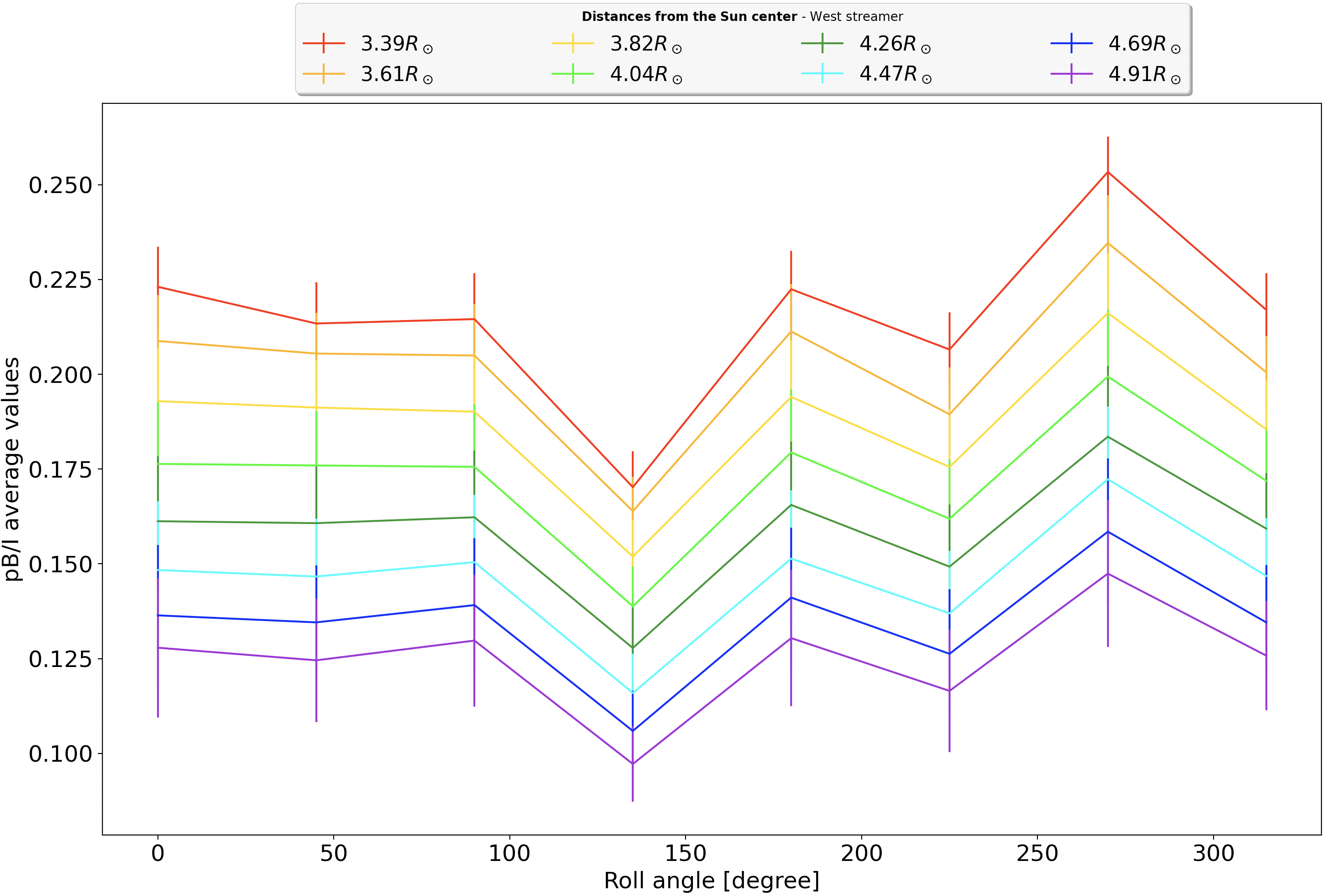}
    \end{tabular}
    \end{center}
    \caption[]{\label{fig:DoLP_Roll_values_Xtheo} 
    $pB$/$I$ average over the pixels inside the selected regions by using the theoretical demodulation tensor. The bars represent the uncertainty inside the considered area.}
\end{figure}

As a comparison, \figurename~\ref{fig:DoLP_Roll_values_Xtheo} shows, as a function of roll angle, the $pB$/$I$ derived from the theoretical demodulation tensor $X^\dag$ given by Eq.~\ref{eq:X_demod_matrix_theo}.
The uncertainty of the $pB$/$I$ mean values using the theoretical $X^\dag$ is from a factor of two to three higher than those using the demodulation tensor from the on-ground calibration. 
\tablename~\ref{tab:DoLP_Roll_values} summarises the $pB$/$I$ values for the two cases.  
 
Data reduction with the theoretical $X^\dag$ would return the signal state of polarization, $pB/I$, with an accuracy of only about 10\% (best case). This would be the dominant uncertainty considering that the one due to the uncertainty of the $pB/I$ is $\approx2$\%, in the considered regions. On the other hand, with the experimental $X^\dag$ the uncertainty in deriving the $pB/I$ values drops to $< 5\%$. This is comparable to the $\approx 3\%$ error of $\approx 1\% - 3\% $ due to the data uncertainty. In conclusion, the experimental demodulation tensor from ground calibration returns polarization measurements with an accuracy of $< 5\%$. As estimated in \citep{S_Fineschi_2005}, with an accuracy of $\approx 1\%$ on the pB (to be compared with the obtained 1-3\% due to data uncertainty), we can expect an $1\%$ error on the electron density evaluation.

The retardances calculated from the fitting of the Malus's curves acquired during the spacecraft rolls are: ($46\degree \pm 4\degree$, $82\degree \pm 4\degree$, $130\degree \pm 5\degree$, $176\degree \pm 5\degree$).


\section{Metis first light pB comparison} 
\begin{figure} [ht!]
    \begin{center}
    \begin{tabular}{c}
        \includegraphics[width = 0.80\linewidth]{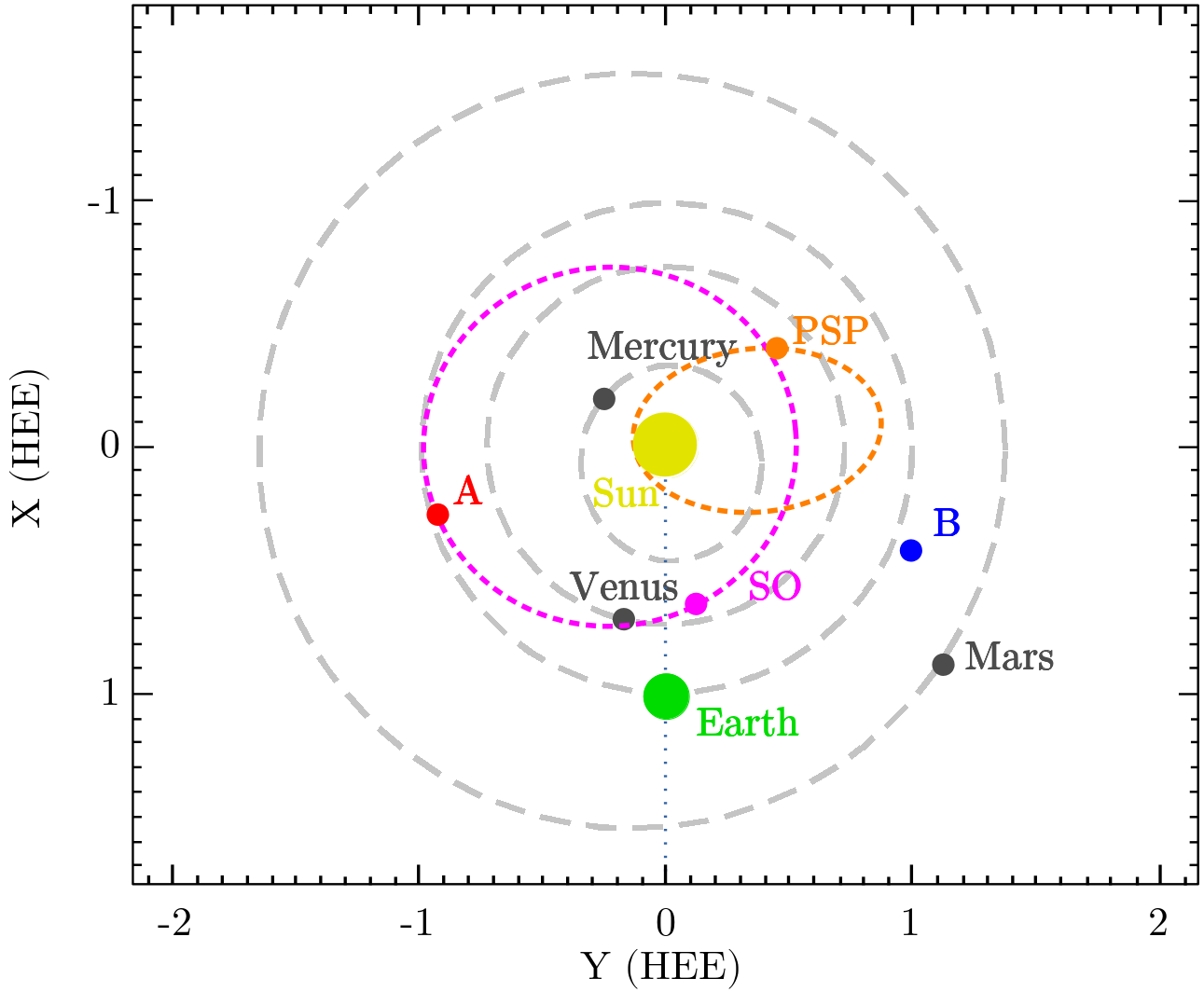}
    \end{tabular}
    \end{center}
    \caption{Position of SolO (SO), STEREO-A (A) and Parker Solar Probe (PSP) with respect to Earth, on May 15, 2020 (00:00 UT). From STEREO webpage.}
    \label{fig:WhereIsSolOandOther}
\end{figure}

\begin{figure} [ht!]
    \subfigure[East streamer]{
      \includegraphics[height = 0.75\linewidth]{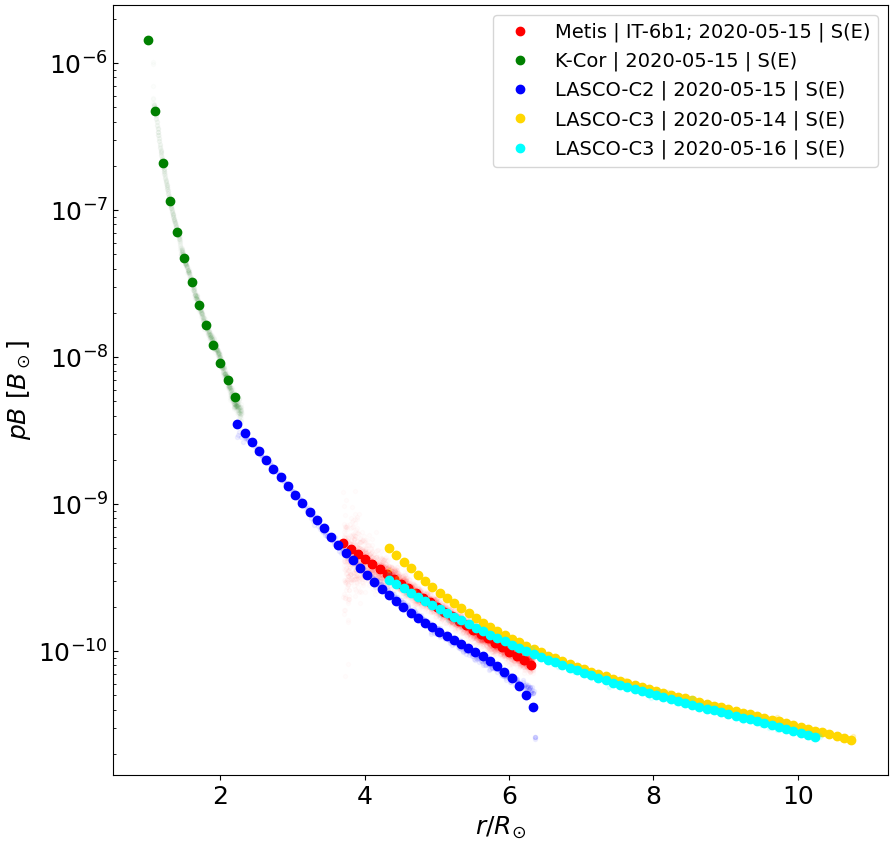}
      \label{fig:pB_Metis_2020May15_and_comparison_a}
    }
     \subfigure[West streamer]{
      \includegraphics[height = 0.75\linewidth]{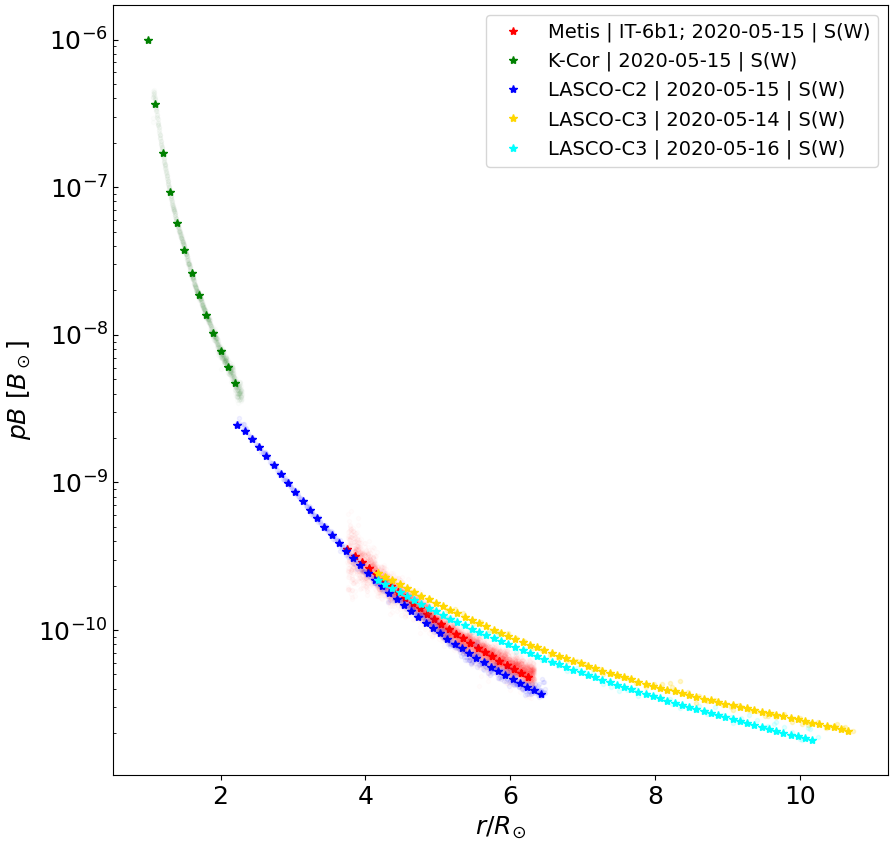}
      \label{fig:pB_Metis_2020May15_and_comparison_b}
    }
    \caption[]{\label{fig:pB_Metis_2020May15_and_comparison} 
    Comparison of $pB$ values along the East and West streamer between Metis, LASCO C2/C3 and K-Cor instruments during an almost Sun-S/C-Earth alignment (data: 2020-05-15; Sun-S/C distance: 0.64\unit{A.U.}).\citep{Fineschi_2021_SPIE}}
\end{figure}

A preliminary verification of the in-flight validation was obtained from the comparison of the coronal polarized brightness as measured by Metis with that obtained by other instruments. In particular, we compared the Metis $pB$ with that of the K-Cor \citep{Hou_2013} and LASCO C2/C3 coronagraphs \citep{Brueckner_1995}. We performed this comparison by using the data of May 15$^{th}$, 2020. At this time Solar Orbiter was almost along the Sun-Earth axis and for this reason LASCO C2 and C3, K-Cor and Metis could be considered aligned (\figurename~\ref{fig:WhereIsSolOandOther}). LASCO-C3 data were available only on the 14$^{th}$ and 16$^{th}$ of May, 2020.
\figurename~\ref{fig:pB_Metis_2020May15_and_comparison}, shows the Metis $pB$ value compared to the LASCO and KCor measurements. On the East streamer Metis $pB$ values are in agreement with those of LASCO-C3, while it is less than a factor of two higher than that of LASCO-C2. On the other hand, on the West limb, Metis and LASCO-C2 measured values agree better.
As the Solar Orbiter mission will progress, more Metis $pB$ measurements will allow a refinement of the cross-calibration with other instruments. These initial measurements were a first check of the performances of the Metis polarimeter, and they succeeded in confirming the validity of its on-ground and in-flight validation. \citep{Fineschi_2021_SPIE}.

\section{Conclusion}
\label{sec:conclusion}  
After a summary of the Metis on-ground calibration, we report its first in-flight validation. 

The first step was the validation of the LCVR retardances. We considered the geometry of the physical process that polarizes the K-Corona radiation with a polarization vector tangent to the solar limb. At four LCVR retardances values, we acquired four images from which we obtain four Malus curves. From these curves we were able to derives the effective LCVR retardances. We obtain retardances, that is, \say{polarization angles} of: ($45.0\degree \pm 0.1\degree$, $81.4\degree \pm 0.1\degree$, $128.7\degree \pm 0.1\degree$, $175.4\degree \pm 0.1\degree$). We also considered a fixed region in the solar corona during a complete S/C rotation as ulterior check. This gives the same result as \say{rotating the polarizer}. We obtain: ($46\degree \pm 4\degree$, $82\degree \pm 4\degree$, $130\degree \pm 5\degree$, $176\degree \pm 5\degree$) that are consistent with the on ground calibration and with what we obtained by considering the polarization of the K-Corona tangent to the solar limb.

We acquired polarimetric sequences with different quadruplets of the used LCVR voltages to evaluate the polarimetric response at different voltage configurations. We considered four fixed coronal regions and obtained the polarized brightness by using the demodulation tensor associated with each quadruplet. This $pB$ comparison from quadruplets with different applied tensions to the polarimeter yielded differences within $\le 2\%$. 

We also evaluated the polarimetric flat field by considering a given coronal regions as it was imaged across different locations of the detector focal plane, during a complete S/C roll. In particular, we considered different heliocentric heights in the $pB/I$ images. As expected, these values have a small percentage variation (i.e., $< 5\%$) at the same heliocentric height during the roll manoeuvre. The percentage variation of the $pB/I$ mean values using the theoretical demodulation tensor is higher than the one obtained using the on ground calibrations. Quantitatively, by using the demodulation tensor from ground calibration, the accuracy in the polarization measurement improves by a factor of $\approx 3$. Indeed, the accuracy of the $pB/I$ values by using the theoretical $X^\dag$ is about 10\% while, by using the $X^\dag$ obtained during on ground calibration, we have an accuracy of $3-5\%$ in the $pB/I$ values. 

Finally, a comparison of Metis polarized brightness with what obtained by other instruments (LASCO C2 and C3  K-Cor) confirms the goodness of the Metis polarimeter calibration.

\vspace{6mm}
\begin{acknowledgements}
\label{end:acknowls} 
This paper has been possible thanks to the whole Metis team and the contributions of the many listed authors. Among the others, a particular acknowledgment to E. Antonucci, who led this Project as Principal Investigator until the delivery of the instrument to ESA in September 2017, when M. Romoli took over this role.

This work has been supported by the Italian Space Agency - ASI. 
The industrial contractor for the Metis project has been a temporary consortium between OHB Italia (for opto-mechanical design, electronics and the software) and Thales Alenia Space Italia (for telescope thermal and structural design and realization, application software, instrument integration, alignment and test). The primary and secondary mirrors were provided as Czech contribution to Metis; the mirror hardware development and manufacturing was possible thanks to the Czech PRODEX Programme. The authors thank also ALTEC Company for providing logistic and technical support. 
\end{acknowledgements}

\bibliographystyle{aa}  
\bibliography{report.bib}   

\begin{thebibliography}{17}
\expandafter\ifx\csname natexlab\endcsname\relax\def\natexlab#1{#1}\fi

\bibitem[{Alvarez-Herrero {et~al.}(2011)Alvarez-Herrero, Uribe-Patarroyo,
  Parejo, Vargas, Heredero, Restrepo, Martínez-Pillet, del Toro~Iniesta,
  López, Fineschi, Capobianco, Georges, López, Boer, \&
  Manolis}]{A_AlvarezHerrero_2011}
Alvarez-Herrero, A., Uribe-Patarroyo, N., Parejo, P.~G., {et~al.} 2011, in
  Polarization Science and Remote Sensing V, ed. J.~A. Shaw \& J.~S. Tyo, Vol.
  8160, International Society for Optics and Photonics (SPIE), 312 -- 329

\bibitem[{Antonucci {et~al.}(2020)Antonucci, Romoli, Andretta, Fineschi,
  Heinzel, Moses, Naletto, Nicolini, Spadaro, Teriaca, \&
  et~al.}]{E_Antonucci_2020}
Antonucci, E., Romoli, M., Andretta, V., {et~al.} 2020, Astronomy \&
  Astrophysics, 642, A10

\bibitem[{{Brueckner} {et~al.}(1995){Brueckner}, {Howard}, {Koomen},
  {Korendyke}, {Michels}, {Moses}, {Socker}, {Dere}, {Lamy}, {Llebaria},
  {Bout}, {Schwenn}, {Simnett}, {Bedford}, \& {Eyles}}]{Brueckner_1995}
{Brueckner}, G.~E., {Howard}, R.~A., {Koomen}, M.~J., {et~al.} 1995, \solphys,
  162, 357

\bibitem[{Casti {et~al.}(2018)Casti, Fineschi, \& et~al.}]{M_Casti_2018}
Casti, M., Fineschi, S., \& et~al. 2018, in {Calibration of the liquid crystal
  visible-light polarimeter for the Metis/Solar Orbiter coronagraph}, ed.
  M.~Lystrup, H.~A. MacEwen, G.~G. Fazio, N.~Batalha, N.~Siegler, \& E.~C.
  Tong, Vol. 10698, International Society for Optics and Photonics (SPIE), 930
  -- 943

\bibitem[{Casti {et~al.}(2019)Casti, Fineschi, \& et~al.}]{M_Casti_2019}
Casti, M., Fineschi, S., \& et~al. 2019, in International Conference on Space
  Optics — ICSO 2018, ed. Z.~Sodnik, N.~Karafolas, \& B.~Cugny, Vol. 11180,
  International Society for Optics and Photonics (SPIE), 1255 -- 1269

\bibitem[{{Elmore} {et~al.}(2000){Elmore}, {Card}, {Lecinski}, {Lites},
  {Streander}, \& {Tomczyk}}]{D_Elmore_2000}
{Elmore}, D.~F., {Card}, G.~L., {Lecinski}, A.~R., {et~al.} 2000, in Society of
  Photo-Optical Instrumentation Engineers (SPIE) Conference Series, Vol. 4139,
  {Calibration procedure for the polarimetric instrument for Solar Eclipse-98},
  ed. S.~{Fineschi}, C.~M. {Korendyke}, O.~H. {Siegmund}, \& B.~E. {Woodgate},
  370--377

\bibitem[{Fineschi {et~al.}(2020)Fineschi, Naletto, Romoli, \&
  et~al.}]{S_Fineschi_2020}
Fineschi, S., Naletto, G., Romoli, M., \& et~al. 2020, Experimental Astronomy,
  49

\bibitem[{{Fineschi} {et~al.}(2021){Fineschi}, {Romoli}, {Andretta},
  {Bemporad}, {Capobianco}, {Casti}, {Da Deppo}, {De Leo}, {Fabi}, \&
  {Frassetto}}]{Fineschi_2021_SPIE}
{Fineschi}, S., {Romoli}, M., {Andretta}, V., {et~al.} 2021, in Society of
  Photo-Optical Instrumentation Engineers (SPIE) Conference Series, Vol. 11852,
  Society of Photo-Optical Instrumentation Engineers (SPIE) Conference Series,
  1185211

\bibitem[{{Fineschi} {et~al.}(2005){Fineschi}, {Zangrilli}, {Rossi}, {Gori},
  {Romoli}, {Corti}, {Capobianco}, {Antonucci}, \& {Pace}}]{S_Fineschi_2005}
{Fineschi}, S., {Zangrilli}, L., {Rossi}, G., {et~al.} 2005, in Society of
  Photo-Optical Instrumentation Engineers (SPIE) Conference Series, Vol. 5901,
  Solar Physics and Space Weather Instrumentation, ed. S.~{Fineschi} \& R.~A.
  {Viereck}, 389--399

\bibitem[{{Hou} {et~al.}(2013){Hou}, {de Wijn}, \& {Tomczyk}}]{Hou_2013}
{Hou}, J., {de Wijn}, A.~G., \& {Tomczyk}, S. 2013, \apj, 774, 85

\bibitem[{Inhester(2016)}]{B_Inhester_2016}
Inhester, B. 2016, arXiv: Solar and Stellar Astrophysics, \textnormal{v2}, 104

\bibitem[{Liberatore {et~al.}(2021)Liberatore, Fineschi, Casti, Capobianco,
  Romoli, Andretta, Bemporad, Deppo, Leo, Fabi, Frassetto, Grimani, Heerlein,
  Heinzel, Jerse, Landini, Magli, Naletto, Nicolini, Pancrazzi, Pelizzo,
  Romano, Sasso, Slemer, Spadaro, Straus, Susino, Teriaca, Uslenghi,
  Volpicelli, \& Zuppella}]{Liberatore_2021_SPIE}
Liberatore, A., Fineschi, S., Casti, M., {et~al.} 2021, in International
  Conference on Space Optics — ICSO 2020, ed. B.~Cugny, Z.~Sodnik, \&
  N.~Karafolas, Vol. 11852, International Society for Optics and Photonics
  (SPIE), 1793 -- 1814

\bibitem[{Müller {et~al.}(2020)Müller, St.~Cyr, Zouganelis, Gilbert, Marsden,
  Nieves-Chinchilla, Antonucci, Auchère, Berghmans, Horbury, \&
  et~al.}]{M_ller_2020}
Müller, D., St.~Cyr, O.~C., Zouganelis, I., {et~al.} 2020, Astronomy \&
  Astrophysics, 642, A1

\bibitem[{{Raouafi}(2011)}]{NE_Raouafi_2011}
{Raouafi}, N.~E. 2011, in Astronomical Society of the Pacific Conference
  Series, Vol. 437, Solar Polarization 6, ed. J.~R. {Kuhn}, D.~M. {Harrington},
  H.~{Lin}, S.~V. {Berdyugina}, J.~{Trujillo-Bueno}, S.~L. {Keil}, \&
  T.~{Rimmele}, 99

\bibitem[{{Solanki} {et~al.}(2020){Solanki}, {del Toro Iniesta}, {Woch},
  {Gandorfer}, {Hirzberger}, {Alvarez-Herrero}, {Appourchaux}, {Mart{\'\i}nez
  Pillet}, {P{\'e}rez-Grande}, {Sanchis Kilders}, {Schmidt}, {G{\'o}mez Cama},
  {Michalik}, {Deutsch}, {Fernandez-Rico}, {Grauf}, {Gizon}, {Heerlein},
  {Kolleck}, {Lagg}, {Meller}, {M{\"u}ller}, {Sch{\"u}hle}, {Staub}, {Albert},
  {Alvarez Copano}, {Beckmann}, {Bischoff}, {Busse}, {Enge}, {Frahm},
  {Germerott}, {Guerrero}, {L{\"o}ptien}, {Meierdierks}, {Oberdorfer},
  {Papagiannaki}, {Ramanath}, {Schou}, {Werner}, {Yang}, {Zerr}, {Bergmann},
  {Bochmann}, {Heinrichs}, {Meyer}, {Monecke}, {M{\"u}ller}, {Sperling},
  {{\'A}lvarez Garc{\'\i}a}, {Aparicio}, {Balaguer Jim{\'e}nez}, {Bellot
  Rubio}, {Cobos Carracosa}, {Girela}, {Hern{\'a}ndez Exp{\'o}sito}, {Herranz},
  {Labrousse}, {L{\'o}pez Jim{\'e}nez}, {Orozco Su{\'a}rez}, {Ramos},
  {Barandiar{\'a}n}, {Bastide}, {Campuzano}, {Cebollero}, {D{\'a}vila},
  {Fern{\'a}ndez-Medina}, {Garc{\'\i}a Parejo}, {Garranzo-Garc{\'\i}a},
  {Laguna}, {Mart{\'\i}n}, {Navarro}, {N{\'u}{\~n}ez Peral}, {Royo},
  {S{\'a}nchez}, {Silva-L{\'o}pez}, {Vera}, {Villanueva}, {Fourmond}, {de
  Galarreta}, {Bouzit}, {Hervier}, {Le Clec'h}, {Szwec}, {Chaigneau},
  {Buttice}, {Dominguez-Tagle}, {Philippon}, {Boumier}, {Le Cocguen},
  {Baranjuk}, {Bell}, {Berkefeld}, {Baumgartner}, {Heidecke}, {Maue}, {Nakai},
  {Scheiffelen}, {Sigwarth}, {Soltau}, {Volkmer}, {Blanco Rodr{\'\i}guez},
  {Domingo}, {Ferreres Sabater}, {Gasent Blesa}, {Rodr{\'\i}guez
  Mart{\'\i}nez}, {Osorno Caudel}, {Bosch}, {Casas}, {Carmona}, {Herms},
  {Roma}, {Alonso}, {G{\'o}mez-Sanjuan}, {Piqueras}, {Torralbo}, {Fiethe},
  {Guan}, {Lange}, {Michel}, {Bonet}, {Fahmy}, {M{\"u}ller}, \&
  {Zouganelis}}]{Solanki_2020}
{Solanki}, S.~K., {del Toro Iniesta}, J.~C., {Woch}, J., {et~al.} 2020, \aap,
  642, A11

\bibitem[{{Van De Hulst}(1950)}]{HC_VanDeHulst_1950}
{Van De Hulst}, H.~C. 1950, B.A.N., 11, 135

\bibitem[{{Zangrilli} {et~al.}(2009){Zangrilli}, {Fineschi}, \&
  {Capobianco}}]{L_Zangrilli_2009}
{Zangrilli}, L., {Fineschi}, S., \& {Capobianco}, G. 2009, in Society of
  Photo-Optical Instrumentation Engineers (SPIE) Conference Series, Vol. 7438,
  Solar Physics and Space Weather Instrumentation III, ed. S.~{Fineschi} \&
  J.~A. {Fennelly}, 74380W

\end{thebibliography}

\clearpage

\begin{table}[ht]
  \caption[Retardances of the different LCVR quadruplets]{Retardances of the different LCVR quadruplets \say{Q} \citep{Liberatore_2021_SPIE}.} 
    \label{tab:quadruplets_details}
    \begin{center}       
        \begin{tabular}{lccc}
        \noalign{\smallskip}\hline
        \rule[-1ex]{0pt}{3.5ex} 
           &  \textbf{Effective angle, $\rho_i$\ [\textdegree]}  &  \\
        \textbf{Q}   &  (difference between \textit{input} and & \textbf{Applied voltage} [mV] \\
           &  \textit{output} polarization directions)  &  \\        
           
        \noalign{\smallskip}\hline\noalign{\smallskip}

        \rule[-1ex]{0pt}{3.5ex} 
        1  &  17.0\textdegree, 60.7\textdegree, 108.9\textdegree, 157.6\textdegree & 30583, 13216, 8344, 6597 \\
        
        \rule[-1ex]{0pt}{3.5ex} 
        2  &  24.6\textdegree, 72.4\textdegree, 121.1\textdegree, 169.4\textdegree & 25362, 11359, 7776, 6313 \\

        \rule[-1ex]{0pt}{3.5ex} 
        3  &  37.3\textdegree, 96.4\textdegree, 145.5\textdegree, 194.6\textdegree & 19573, 9087, 6924, 5810 \\
        
        \rule[-1ex]{0pt}{3.5ex} 
        4 (nominal)  &  49.1\textdegree, 84.3\textdegree, 133.2\textdegree, 181.1\textdegree & 15837, 10048, 7318, 6051 \\
        \noalign{\smallskip}\hline
        \end{tabular}
    \end{center}
\end{table} 

\begin{table}[ht]
    \caption[]{$pB$/$I$ average values with standard deviation ($pB$ and $I$ obtained through the on-ground demodulation tensor [left] and theoretical one [right]). Uncertainty column contains the standard deviation obtained considering the $pB$/$I$ values in the region at fixed heliocentric distance (white squares in \figurename~\ref{fig:DoLP_Roll}) for each roll.} 
    \label{tab:DoLP_Roll_values}
    \begin{center}       
    \begin{tabular}{c|cc|cc|cc|}

    \rule[-1ex]{0pt}{3.5ex} 
      &  \multicolumn{2}{c|}{On ground $X^\dag$ ($4.26\unit{R_\odot}$)} &  \multicolumn{2}{c}{On ground $X^\dag$ ($3.39\unit{R_\odot}$)} & \multicolumn{2}{|c|}{Theoretical $X^\dag$ ($4.26\unit{R_\odot}$)} \\
      &  \multicolumn{2}{c|}{Best case} &  \multicolumn{2}{c}{Worst case} & \multicolumn{2}{|c|}{Best case} \\
    \noalign{\smallskip}\hline\noalign{\smallskip}

    \rule[-1ex]{0pt}{3.5ex} 
    Roll & pB/I & $\sigma$ & pB/I & $\sigma$ & pB/I & $\sigma$ \\

    \noalign{\smallskip}\hline\noalign{\smallskip}

    \rule[-1ex]{0pt}{3.5ex} 
    0 & 0.175 & 0.018 & 0.253 & 0.027 & 0.161 & 0.017\\

    \rule[-1ex]{0pt}{3.5ex} 
    1 & 0.173 & 0.018 & 0.246 & 0.036 & 0.160 & 0.015\\

    \rule[-1ex]{0pt}{3.5ex} 
    2 & 0.172 & 0.019 & 0.231 & 0.023 & 0.162 & 0.017\\

    \rule[-1ex]{0pt}{3.5ex} 
    3 & 0.175 & 0.021 & 0.227 & 0.031 & 0.128 & 0.010\\

    \rule[-1ex]{0pt}{3.5ex} 
    4 & 0.176 & 0.019 & 0.236 & 0.018 & 0.166 & 0.017\\

    \rule[-1ex]{0pt}{3.5ex} 
    5 & 0.173 & 0.018 & 0.239 & 0.026 & 0.149 & 0.016\\

    \rule[-1ex]{0pt}{3.5ex} 
    6 & 0.188 & 0.019 & 0.268 & 0.022 & 0.184 & 0.019\\
    
    \rule[-1ex]{0pt}{3.5ex} 
    7 & 0.180 & 0.018 & 0.245 & 0.040 & 0.159 & 0.014\\

    \noalign{\smallskip}\hline\noalign{\smallskip}

    \rule[-1ex]{0pt}{3.5ex} 
    Avg. & 0.177 & 0.019 & 0.243 & 0.028 & 0.159 & 0.016 \\
    \rule[-1ex]{0pt}{3.5ex} 
    StD. & 0.005 & 0.001 & 0.013 & 0.008 & 0.016 & 0.002 \\
    
    \noalign{\smallskip}\hline\noalign{\smallskip}
    
    \rule[-1ex]{0pt}{3.5ex} 
    StD/Avg & 3\% & 1\% & 5\% & 3\% & 10\% & 2\% \\

     \noalign{\smallskip}\hline
    
    \end{tabular}
    \end{center}
\end{table}

\end{document}